\documentstyle[12pt]{article}
\def\be {\begin{equation}}
\def\ee {\end{equation}}
\def\ba {\begin{eqnarray}}
\def\ea {\end{eqnarray}}

\def\bi {\begin{itemize}}
\def\ei {\end{itemize}}
\begin{document}
\def\bea{\begin{eqnarray}}
\def\eea{\end{eqnarray}}
\title{\bf {Entropy of Scalar Field near a Schwarzschild Black Hole Horizon }}
 \author{M.R. Setare  \footnote{E-mail: rezakord@ipm.ir}
  \\{Department of Science,  Payame Noor University. Bijar, Iran}}
\date{\small{}}

\maketitle
\begin{abstract}
In this paper we compute the correction to the entropy of
Schwarzschild black hole due to the vacuum polarization effect of
massive scalar field. The Schwarzschild black hole is supposed to be
confined in spherical shell. The scalar field obeying mixed boundary
condition on the spherical shell.
 \end{abstract}
\newpage

 \section{Introduction}
In general relativity, black hole's properties can be precisely
calculated, and the holes may be thought of astronomical objects
with masses about several times of sun. In this classical case,
event horizon emerges, and anything can not escape from it to arrive
at a particular observer who is outside the horizon. But Hawking
found that black hole emits radiation and the radiation spectra is
just the black body's. This fact indicates that the black hole has a
temperature whose expression is \cite{{how},{how1},{how2}}
\begin{equation}\label{hawte}
T_{BH}=\frac{k}{2\pi},
\end{equation}
where $k$ is the surface gravity on the horizon. Thus the
Bekenstein-relation becomes the real thermodynamical relationship
of the black hole. Namely the black hole has a thermodynamical
entropy as
\begin{equation}\label{hawen}
S_{BH}=\frac{A}{4l_{p}^{2}}
\end{equation}
where $S_{BH}$ is the Bekenstein-Hawking entropy, $A$ is the area
of the event horizon and $l_p=(\frac{\hbar G}{c^3})^{1/2}$ is the
Planck length. The presence of quantum field in black hole
background modifies the entropy. In the state of thermal
equilibrium the total entropy is as \cite{zak}
\begin{equation}\label{sto}
 S_{tot}=S_{BH}+S_q
\end{equation}
where $S_q$ is the contribution of radiation and matter fields.
Quantum correction to the Bekenstein-Hawking entropy,due to a
scalar field, have been computed by different methods for
Schwarzschild\cite{cog, fro}and Reissner-Nordstrom
{\cite{dem}-\cite{rome}} black holes.\\
A very simple but nonetheless instructive model to address the
problem of black hole entropy is the so called ``brick wall'' by
't Hooft \cite{tho}. 't Hooft considers a quantum field in the
background of a classical black hole and using the WKB
approximation he derives the thermal entropy of the field outside
the horizon of the black hole. In performing the computation, two
spatial cutoffs are employed: a large distance one, needed to
avoid large volume divergences in the asymptotically flat region,
and a short distance one, the ``brick wall'', localized just
outside the horizon and suppressing the divergences due to the
growing number of modes close to the horizon. On the boundaries of
the space slice arising in this way, Dirichlet boundary conditions
are imposed. The entropy obtained in this model is divergent in
the limit of vanishing brick wall thickness. These divergences
were later recognized as quantum corrections to the
Bekenstein-Hawking formula which can be absorbed into
renormalization of the one loop effective gravitational lagrangian
\cite{Susskind:1994sm,Fursaev:1994ea, winstanley, Brustein:2005vx, mila} \\
In this letter we would like to investigate the case of
Schwarzschild black hole. We consider the massive scalar field
obeying mixed boundary condition on a spherical shell, the
Schwarzschild black hole is supposed to be confined in the
spherical container, in other words we revisit the brick wall
model here. Then we obtain the contribution $S_q$ of scalar field
onto entropy of a black hole.
$S_q$ is due to vacuum polarization effect.\\
In the generic black hole background the investigation of
boundary-induced quantum effects is techniqally complicated, the
exact analytical result can be obtained in the near horizon and
large mass limit when the boundary is close to the black hole
horizon. In this limit the black hole geometry may be approximated
by the Rindler-like manifold (for some investigations of quantum
effects on background of Rindler-like spactimes see
\cite{{bys},{zer},{zer1}} ). We consider the vacuum expectation
values of the field square and the energy-momentum tensor for a
conformally coupled scalar field in the presence of spherical shell
on the bulk $Ri\times S^2$, where $Ri$ is a two-dimensional Rindler
spacetime.
\section{Massive scalar field in Rindler-like spacetime}
Let us consider a scalar field $\varphi (x)$ at finite
temperature equal to its Hawking value $T=\beta^{-1}$,
propagating on background of four-dimensional Rindler-like
spacetime $Ri\times S^{2}$, where $Ri$ is a two-dimensional
Rindler spacetime. The corresponding metric is described
by the line element%
\begin{equation}
ds^{2}=\xi ^{2}d\tau ^{2}-d\xi ^{2}-r_{H}^{2}d\Sigma _{2}^{2},
\label{ds22}
\end{equation}%
with the Rindler-like $(\tau ,\xi )$ part and $d\Sigma _{2}^{2}$
is the line element for the space with positive constant curvature
with the Ricci scalar $R=2/r_{H}^{2}$. Line element (\ref{ds22})
describes the near horizon geometry of four-dimensional AdS black
hole with the line element \cite{{ds22},{van}}%
\begin{equation}
ds^{2}=A_{H}(r)dt^{2}-\frac{dr^{2}}{A_{H}(r)}-r^{2}d\Sigma
_{2}^{2}, \label{ds21}
\end{equation}%
where%
\begin{equation}
A_{H}(r)=k+\frac{r^{2}}{l^{2}}-\frac{r_{0}^{3}}{l^{2}r},
\label{Ar}
\end{equation}%
and $k=0,-1,1$. In (\ref{Ar}) the parameter $l$ is related to the
bulk cosmological constant and the parameter $r_0$ depends on the
mass of the black hole and on the bulk gravitational constant. In
the non extremal case the function $A_H(r)$ has a simple zero at
$r=r_H$. In the near horizon limit, introducing new coordinates
$\tau $ and $\rho $
in accordance with%
\begin{equation}
\tau =\frac{1}{2}A_{H}^{\prime }(r_{H})t,\quad r-r_{H}=\frac{1}{4}%
A_{H}^{\prime }(r_{H})\xi ^{2},  \label{tau}
\end{equation}%
the line element is written in the form (\ref{ds22}). Note that
for a $(3+1)$-dimensional Schwarzschild black hole one has
$A_H(r)=1-\frac{r_H}{r}$ and, hence, $A'_H(r_H)=1/r_{H}$. The
field equation is in the form
\begin{equation}
\left( g^{ik}\nabla _{i}\nabla _{k}+m^{2}+\zeta R\right) \varphi
(x)=0, \label{fieldeq1}
\end{equation}%
where $\zeta $ is the curvature coupling parameter. Below we will assume that the field satisfies
the Robin boundary condition%
\begin{equation}
\left( A+B\frac{\partial }{\partial \xi }\right) \varphi =0,\quad
\xi =a, \label{bound1}
\end{equation}%
on the hypersurface $\xi =a$, with constant coefficients $A$ and
$B$.  In accordance with \ (\ref{tau}) this hypersurface
corresponds to the spherical shell near the black hole horizon
with the radius $r_{a}=r_{H}+A_{H}^{\prime }(r_{H})a^{2}/4$.\\
A black hole behaves like a thermodynamic system and possesses
temperature and entropy. As we have mentioned in introduction in
the state of thermal equilibrium the total entropy is as
Eq.(\ref{sto}). The Euclidean action for scalar filed in our
interesting background takes the standard thermodynamic form
\begin{equation}\label{act}
I_q=-\beta\int d^{3}x\sqrt{g}\left\langle 0_0\left\vert
T_{0}^{0}\right\vert 0_0\right\rangle-S_q
\end{equation}
where $T_{\mu}^{\nu}$ is the stress tensor of the quantum field
calculated with respect to the background (\ref{ds21}), $g$ is
determinant of the metric. From the conservation law
$T^{\mu}_{\nu ;\mu}=0$ in the background (\ref{ds21}) it follows
that \cite{zasla}
\begin{equation}\label{coneq}
\frac{r}{2}\frac{(<T^{1}_{1}>-<T^{0}_{0}>)A'_{H}}{A_H}=\frac{1}{r_H}[(r^3<T^{1}_{1}>)_{,1}-r^2<T^{i}_{i}>]
\end{equation}
where $i$ denotes spatial indices. Then one can obtain (for more
details see \cite{zasla})
\begin{equation}\label{acem1}
(4\pi)^{-2}A'_H(r_H)\frac{\partial I_q}{\partial
r_H}=\int_{r_H}^{r_a}dr
r^2<T^{i}_{i}>-r_{a}^{3}<T^{1}_{1}>+r_{H}^{3}[<T^{1}_{1}(r_{H})>-<T^{0}_{0}(r_{H})>]
\end{equation}
The regularity of $<T^{\nu}_{\mu}>$ at the event horizon in the
Hartle-Hawking state demands that according to (\ref{coneq})
$<T^{1}_{1}(r_{H})>=<T^{0}_{0}(r_{H})>$, so the last term in
(\ref{acem1}) is square brackets cancels. The above formula are
applicable to any field with a tensor $<T^{\nu}_{\mu}>$.
 Now we consider the Schwarzschild
background case, although one can consider the general line
element (\ref{ds21}), the Schwarzzschild is only a simple case,
in this case we have
\begin{equation}\label{acem}
(4\pi)^{-2}\frac{\partial I_q}{\partial
r_H}=r_{a}^{3}<T^{r}_{r}>-\int_{r_H}^{r_a}dr r^2<T^{i}_{i}>
\end{equation}
\section{Vacuum expectation values of energy-momentum tensor and entropy of scalar field}
Now we consider the vacuum expectation values (VEV) of the
energy-momentum tensor for the geometry $R^{2}\times S^{2}$
described by the following line element
\begin{equation}
ds^{2}=dt^{2}-(dx^{1})^{2}-r_{H}^{2}d\Sigma _{2}^{2}, \label{ds31}
\end{equation}%
where the coordinates $(t,x^{1})$ are related to the coordinates
$(\tau ,\xi )$ by $t=\xi \sinh \tau $, $x^{1}=\xi \cosh \tau $.
This line element describes the near horizon geometry of a
non-extremal black hole spacetime. We will consider the vacuum
expectation values for the geometry without boundaries. The
corresponding Wightman function can be presented in the form
\cite{setsah}
\begin{eqnarray}
G_{0}^{+}(x,x^{\prime }) &=&\tilde{G}_{0}^{+}(x,x^{\prime
})-\frac{\Gamma(3/2)}{2\pi ^{7/2}r_{H}^{2}}\sum_{l=0}^{\infty
}(2l+1)C_{l}^{1/2}(\cos \theta )
\nonumber \\
&&\times \int_{0}^{\infty }d\omega e^{-\omega \pi }\cos [\omega
(\tau -\tau ^{\prime })]K_{i\omega }(\lambda _{l}\xi )K_{i\omega
}(\lambda _{l}\xi ^{\prime }),  \label{GM1}
\end{eqnarray}%
with the function%
\begin{eqnarray}
\tilde{G}_{0}^{+}(x,x^{\prime }) &=&\frac{\Gamma(3/2)}{2\pi ^{7/2}r_{H}^{2}}%
\sum_{l=0}^{\infty }(2l+1)C_{l}^{1/2}(\cos \theta )  \nonumber \\
&&\times \int_{0}^{\infty }d\omega \cosh \{\omega \lbrack \pi
-i(\tau -\tau ^{\prime })]\}K_{i\omega }(\lambda _{l}\xi
)K_{i\omega }(\lambda _{l}\xi ^{\prime }).  \label{GM2}
\end{eqnarray}%
where $K_{i\omega}(\lambda_{l}\xi)$, is the Bessel modified
function with the imaginary order, and $C_{l}^{1/2}(\cos \theta
)$ is the Gegenbauer or ultraspherical polynomial of degree $l$
and order $1/2$. In (\ref{GM1}), the divergences in the
coincidence limit are contained in the term
$\tilde{G}_{0}^{+}(x,x^{\prime })$, the amplitude of the corresponding vacuum state we will denote by $|%
\tilde{0}\rangle $. First of all we note that from the problem
symmetry it follows that the expectation values  $\left\langle \tilde{0}%
\left\vert T_{i}^{i}\right\vert \tilde{0}\right\rangle $ do not
depend on
the point of observation and%
\begin{eqnarray}
\left\langle \tilde{0}\left\vert T_{0}^{0}\right\vert
\tilde{0}\right\rangle
&=&\left\langle \tilde{0}\left\vert T_{1}^{1}\right\vert \tilde{0}%
\right\rangle ,  \label{vevEMTM} \\
\left\langle \tilde{0}\left\vert T_{2}^{2}\right\vert
\tilde{0}\right\rangle
&=&\left\langle \tilde{0}\left\vert T_{3}^{3}\right\vert \tilde{0}%
\right\rangle .  \nonumber
\end{eqnarray}%
The component $\left\langle \tilde{0}\left\vert T_{2}^{2}\right\vert \tilde{0%
}\right\rangle $ can be expressed through the energy density by
using the
trace relation%
\begin{equation}
T_{i}^{i}=3(\zeta -\zeta _{c})\nabla _{i}\nabla ^{i}\varphi
^{2}+m^{2}\varphi ^{2}.  \label{TiiM}
\end{equation}%
From this relation it follows that%
\begin{equation}
\left\langle \tilde{0}\left\vert T_{2}^{2}\right\vert
\tilde{0}\right\rangle =\frac{1}{2}\left[ m^{2}\left\langle
\tilde{0}\left\vert \varphi ^{2}\right\vert \tilde{0}\right\rangle
-2\left\langle \tilde{0}\left\vert T_{0}^{0}\right\vert
\tilde{0}\right\rangle \right] .  \label{vevT22M}
\end{equation}%
Hence, it is sufficient to find the renormalized vacuum
expectation values of the field square and the energy density.
Then as have shown in \cite{setsah}
\begin{equation}\label{phwem}
\left\langle \tilde{0}\left\vert \varphi ^{2}\right\vert \tilde{0}%
\right\rangle =\frac{\zeta(1/2)\Gamma(3/2)}{8\pi^{5/2}r_{H}^{2}},
\hspace{1cm}\left\langle \tilde{0}\left\vert T_{0}^{0}\right\vert
\tilde{0}\right\rangle=\frac{\zeta(-1/2)\Gamma(3/2)}{8\pi^{5/2}r_{H}^{4}}
\end{equation}
Using Eqs.(\ref{vevEMTM},\ref{vevT22M}) we obtain
\begin{equation}\label{phwem2}
\left\langle \tilde{0}\left\vert T_{i}^{i}\right\vert
\tilde{0}\right\rangle=2\left\langle \tilde{0}\left\vert
T_{0}^{0}\right\vert \tilde{0}\right\rangle+2\left\langle
\tilde{0}\left\vert T_{2}^{2}\right\vert
\tilde{0}\right\rangle=m^{2}\left\langle \tilde{0}\left\vert \varphi ^{2}\right\vert \tilde{0}%
\right\rangle
\end{equation}
Now we turn to the VEV of the energy-momentum tensor. The
corresponding
operator we will take in the form%
\begin{equation}
T_{ik}=\partial _{i}\varphi \partial _{k}\varphi +\left[ \left( \zeta -\frac{%
1}{4}\right) g_{ik}\nabla _{l}\nabla ^{l}-\zeta \nabla _{i}\nabla
_{k}-\zeta R_{ik}\right] \varphi ^{2},  \label{EMT1}
\end{equation}%
with the trace relation (\ref{TiiM}). In (\ref{EMT1}) $R_{ik}$ is
the Ricci tensor for the bulk geometry and for the metric
(\ref{ds22}) it has
components%
\begin{eqnarray}
R_{ik} &=&0,\quad i,k=0,1;\quad   \label{Rik} \\
R_{ik} &=&\frac{n}{r_{H}^{2}}g_{ik},\quad i,k=2,3.
\end{eqnarray}%
On the base of formula (\ref{EMT1}) the corresponding vacuum
expectation
values are presented in the form%
\begin{equation}
\langle 0_{0}\left\vert T_{ik}\right\vert 0_{0}\rangle
=\lim_{x^{\prime }\rightarrow x}\nabla _{i}\nabla _{k}^{\prime
}G_{0}^{+}(x,x^{\prime })+ \left[ \left( \zeta -\frac{1}{4}\right)
g_{ik}\nabla _{l}\nabla ^{l}-\zeta \nabla _{i}\nabla _{k}-\zeta
R_{ik}\right] \langle 0_{0}\left\vert \varphi ^{2}\right\vert
0_{0}\rangle .  \label{EMT2}
\end{equation}%
By using decomposition (\ref{GM1}), the vacuum energy-momentum
tensor is
presented in the form%
\begin{equation}
\langle 0_{0}\left\vert T_{ik}\right\vert 0_{0}\rangle =\langle \tilde{0}%
\left\vert T_{ik}\right\vert \tilde{0}\rangle +\langle
T_{ik}\rangle ^{(0)}, \label{Tik0}
\end{equation}%
where the second summand on the right is given by formula
\cite{setsah}
\begin{equation}
\langle T_{i}^{k}(x)\rangle ^{(0)}=\frac{-\delta
_{i}^{k}\Gamma(3/2)}{2\pi ^{7/2}r_{H}^{2}}\sum_{l=0}^{\infty
}D_{l}\lambda _{l}^{2}\int_{0}^{\infty }d\omega e^{-\omega \pi
}f^{(i)}\left[ K_{i\omega }(\lambda _{l}\xi )\right] .
\label{Tik00}
\end{equation}%
In this formula we use the notations%
\begin{eqnarray}
f^{(0)}[g(z)] &=&\left( \frac{1}{2}-2\zeta \right) \left[ \left( \frac{dg(z)%
}{dz}\right) ^{2}+\left( 1-\frac{\omega ^{2}}{z^{2}}\right) g^{2}(z)\right] +%
\frac{\zeta }{z}\frac{d}{dz}g^{2}(z)+\frac{\omega
^{2}}{z^{2}}g^{2}(z),
\label{f(i)} \\
f^{(1)}[g(z)] &=&-\frac{1}{2}\left( \frac{dg(z)}{dz}\right)
^{2}-\frac{\zeta
}{z}\frac{d}{dz}g^{2}(z)+\frac{1}{2}\left( 1-\frac{\omega ^{2}}{z^{2}}%
\right) g^{2}(z),  \nonumber \\
f^{(i)}[g(z)] &=&\left( \frac{1}{2}-2\zeta \right) \left[ \left( \frac{dg(z)%
}{dz}\right) ^{2}+\left( 1-\frac{\omega ^{2}}{z^{2}}\right) g^{2}(z)\right] -%
\frac{\lambda _{l}^{2}-m^{2}}{2\lambda _{l}^{2}}g^{2}(z),
\nonumber
\end{eqnarray}%
also we use the following notations
\begin{equation}
D_l=\frac{(2l+1)\Gamma(l+1)}{l!}
\end{equation}
\begin{equation}
\quad \lambda _{l}=\sqrt{\frac{l(l+1)+2\zeta} {r_{H}^{2}}+m^{2}}.
\label{lambdal}
\end{equation}
 Now by considering Eq.(\ref{acem})we can write
\begin{equation}\label{acem2}
(4\pi)^{-2}\frac{\partial I_q}{\partial
r_H}=r_{a}^{3}\left\langle 0_0\left\vert T_{1}^{1}\right\vert
0_0\right\rangle-\int_{r_H}^{r_a}dr
r^2\left\langle 0_0\left\vert T_{i}^{i}\right\vert 0_0%
\right\rangle
\end{equation}
then using Eqs.(\ref{phwem},\ref{phwem2},\ref{Tik0}) we obtain
\begin{eqnarray}
I_q&=&\frac{\Gamma(3/2)}{6\pi^{^1/2}}[m^2\zeta(1/2)(\frac{r_{a}^{3}}{r_H}+1/2r_{H}^{2})-\frac{\zeta(-1/2)}{r_{H}^3}]+
\frac{r_{a}^{3}}{16\pi^{2}}\int dr_{H}<T^{1}_{1}>^{(0)} \nonumber \\
&-&\frac{1}{16\pi^{2}}\int dr_{H}\int_{r_{H}}^{r_a}dr r^2
<T^{i}_{i}>^{(0)} +c
\end{eqnarray}\label{acti}
where $c$ is a constant. For small values $\xi$ the vacuum
expectation values (\ref{Tik00}) behave as
$(\frac{r_H}{\xi})^{4}$, therefore
\begin{eqnarray}
I_q=\frac{\Gamma(3/2)}{6\pi^{^1/2}}[m^2\zeta(1/2)(\frac{r_{a}^{3}}{r_H}+1/2r_{H}^{2})-\frac{\zeta(-1/2)}{r_{H}^3}]+
\frac{r_{H}^{5}}{48 \pi^{2}\xi^{4}}(r_{a}^{3}/5-r_{H}^{3}/2)+c
\end{eqnarray}\label{acti1}
 Now using Eq.(\ref{act}) we
have
\begin{eqnarray}
S_q&=&\frac{-\beta(r_{a}^{3}-r_{H}^{3})}{6\pi^{3/2}r_{H}^{4}}\zeta(-1/2)\Gamma(3/2)+
\frac{4\pi\beta(r_{a}^{3}-r_{H}^{3})}{3}\frac{r_{H}^{4}}{\xi^{4}}\nonumber \\
&-&
\frac{\Gamma(3/2)}{6\pi^{^1/2}}[m^2\zeta(1/2)(\frac{r_{a}^{3}}{r_H}+1/2r_{H}^{2})-\frac{\zeta(-1/2)}{r_{H}^3}]-
\frac{r_{H}^{5}}{48 \pi^{2}\xi^{4}}(r_{a}^{3}/5-r_{H}^{3}/2)+c
\end{eqnarray}
We have succeded in obtaining a reasonable expression, valid only
for very large black hole mass and near the horizon. On the horizon
the vacuum expectation values (\ref{Tik0},\ref{Tik00}) diverge.
These surface divergences are well known in quantum field theory
with boundaries and are investigated for various type boundary
conditions and geometries. The leading term in the near horizon
asymptotic expansions behaves as $(\frac{r_{H}}{\xi})^{4}$ for the
components of the energy-momentum tensor. On the horizon $\xi=0$ and
$r_a=r_H$, then one can see that the divergent term in the entropy
$S_q$ is proportional to the $\frac{1}{\xi^{4}}$. A possible way to
deal with such divergences has been suggested in \cite{tho}, where
it has been argued that the quantum fluctuations at the horizon
might provide a natural cutoff. In Ref.\cite{tho}, 't Hooft
attempted to provide a microphysical explanation of black hole
entropy by considering the modes for a scalar field in the vicinity
of a black hole. In such a calculation, one finds a divergence in
the number of modes because of the infinite blue shift at the event
horizon. To regulate his calculation, 't Hooft introduced a ``brick
wall'' cut-off, demanding that the scalar field vanish within some
fixed distance of the horizon. 't Hooft introduced this
``simple-minded'' cut-off as an attempt to mimic what he hoped would
be a true physical regulator arising from gravitational
interactions. Susskind and Uglum suggested that the entropy
divergences have the correct form to be absorbed in the
Bekenstein-Hawking formula as a renormalization of Newton's
constant\cite{Susskind:1994sm}. Thus these calculations should be
regarded as yielding the one-loop correction of quantum field theory
to the black hole entropy\cite{callan,tedb,myer}.
\section{Conclusion}
In this paper we studied the quantum vacuum effects to the entropy
produced by a spherical shell in $4-$dimensional $Ri\times S^2$
spacetime, with $Ri$ being a two-dimensional Rindler spacetime. The
corresponding line element (\ref{ds22}) describes the near horizon
geometry of a non-extremal black hole spactime defined by the line
element (\ref{ds21}). The case of a massive scalar field with
conformal coupling parameter and satisfying the Robin boundary
condition on the sphere is considered. Then by considering the
energy-momentum tensor of quantum massive scalar field near the
horizon of Schwarzschild black hole and using general formula
(\ref{acem1}), we recovered the contribution $S_q$ of these field
into the entropy of a black hole.  The vacuum expectation values of
the field square and the energy-momentum tensor are expressed in
terms of the zeta function by formulas (20), at $s=1/2$ and $s=-1/2$
the zeta function $\zeta(s)$ has simple poles with residues
$\zeta_{S^2}(0)$ and $\frac{\zeta_{S^2}(-1)}{2}$, respectively.
Hence, in general, the vacuum expectation values of the field square
and the energy density contain the pole and finite contributions.
The remained pole term is a characteristic feature for the zeta
function regularization method. As a result the vacuum expectation
values of the energy-momentum tensor for the boundary-free geometry
are determined by formulas (26), (27). On the horizon these
expectation values diverge. The leading terms in the near horizon
asymptotic expansion behave as $(\frac{r_{H}}{\xi})^{4}$ for the
components of the energy-momentum tensor.  On the horizon $\xi=0$
and $r_a=r_H$, then one can see that the divergent term in the
entropy $S_q$ is proportional to the $\frac{1}{\xi^{4}}$. These
surface divergences are well known in quantum field theory with
boundaries and are investigated for various type boundary conditions
and geometries. Local surface divergences were first discussed for
arbitrary smooth boundaries by Deutsch and Candelas
\cite{Deutsch:1978sc}.  They found cubic divergences in the energy
density as one approaches the surface; for example, outside a
Dirichlet sphere (that is, for a conformally-coupled scalar field
satisfying Dirichlet boundary conditions on the surface) the energy
density diverges. The first calculations on the problem of
divergences in one--loop thermodynamical quantities for matter
fields in thermal equilibrium on a black hole background are due to
G. `t Hooft \cite{tho}: using a WKB approximation for the
eigenvalues of a scalar field hamiltonian on the Schwarz\-schild
background, `t Hooft finds that thermodynamical quantities as free
energy, internal energy and entropy have contributions divergent for
the radial coordinate $r\rightarrow r_H$. So one must introduce a
short--distance cut--off $\epsilon$ representing a radial proper
distance from the horizon. The divergences in the thermodynamical
quantities behave as
 $\epsilon^{-2}$. `t Hooft proposal to face with these divergences
is the so called brick wall model.
  

\vspace{3mm}








\begin{thebibliography}{99}
\bibitem{how}S. W. Hawking, Nature, {\bf 248}, 30, (1974).
\bibitem{how1}S. W. Hawking, Commun. Math. Phys. {\bf 43}, 199, (1975).
\bibitem{how2} S. W. Hawking, Phys. Rev. {\bf D14}, 2460, (1976).
\bibitem{zak}O. B. Zaslavskii, gr-qc/0101055.
\bibitem{cog}G. Cognola, L. Vanzo and S. Zerbini, Class. Quant.
Grav. {\bf 12}, 1927, (1995).
\bibitem{fro}V. P. Frolov, D. V. Furasev and A. I. Zelnikov, Phys.
Rev. {\bf D54}, 2711, (1996).
\bibitem{dem}J. Demers, R. Lafrance and R. C. Myers, Phys.
Rev.{\bf D52}, 2245, (1995).
\bibitem{gosh}A. Ghosh and P. Mitra, Phys. Lett. {\bf B357}, 295,
(1995).
\bibitem{cog1}G. Cognola, L. Vanzo and S. Zerbini,  Phys.
Rev.{\bf D52}, 4548, (1995).
\bibitem{gos}A. Ghosh and P. Mitra, Mod. Phys. Lett. {\bf A11},
2933, (1996).
\bibitem{ohta}S. P. de Alwis and N. Ohta, hep-th/9504033.
\bibitem{rome}A. Romeo,  Class. Quant.
Grav. {\bf 13}, 2797, (1996).
\bibitem{tho}G.'t Hooft. Nucl. Phys. {\bf B256}, 727 (1985).
\bibitem{Susskind:1994sm}
L.~Susskind and J.~Uglum, Phys. Rev. {\bf D50}, 2700, (1994).

\bibitem{Fursaev:1994ea}
D.~V. Fursaev and S.~N. Solodukhin, Phys. Lett. {\bf B365}, 51,
(1996).
\bibitem{winstanley}
E.~Winstanley, Phys. Rev. {\bf D 63}, 084013, (2001).
\bibitem{Brustein:2005vx}
R.~Brustein, M.~B. Einhorn, and A.~Yarom, hep-th/0508217.
\bibitem{mila}G. Milanesi, M. Mintchev, hep-th/0509080.
\bibitem{bys}A. A. Bytsenko, G. Cognola, and S. Zerbini, Nucl.
Phys. {\bf B458}, 267, (1996).
\bibitem{zer} S. Zerbini, G. Cognola,
and L. Vanzo, Phys. Rev.{\bf D54}, 2699, (1995).
\bibitem{zer1} G.
Cognola, E. Elizalde and S. Zerbini,  Phys. Lett. {\bf B585}, 155,
(2004).
\bibitem{ds22} R. B. Mann,  Class. Quant.
Grav. {\bf 16}, L109, (1997).
\bibitem{van}  L. Vanzo,  Phys. Rev. {\bf D56}, 6475, (1997).
\bibitem{zasla}O. B. Zaslavskii, Class. Quant.
Grav. {\bf 13}, L23, (1996).
\bibitem{setsah} A. A. Saharian, M. R. Setare, Nucl. Phys. {\bf B724}, 406, (2005).
\bibitem{callan} C. Callan and F. Wilczek, {\em Phys. Lett.} {\bf
B333}, 55, (1994).
\bibitem{tedb} T. Jacobson, {\em Phys. Rev.} {\bf D50}, 6031, (1994).

\bibitem{myer}J. G. Demers, R. Lafrance, R. C. Myers, Phys. Rev. {\bf D52}, 2245, (1995).
\bibitem{Deutsch:1978sc}D. Deutsch  and P. Candelas ,
 Phys. Rev.  {\bf D20}, 3063, (1979).
\end{thebibliography}
\end{document}